\documentclass[10pt,fleqn,a4paper,twoside]{article}
\usepackage{eurodiname2026, bm}
\def\shortauthor{G.S. Lima and W.M. Bessa}
\def\shorttitle{Learning-based control of a single-DOF Aero system}

\usepackage{pgfplots}
\usepackage{tikz}
\usepackage{amssymb, amsthm, amsmath}
\newcommand*{\myfont}{\fontfamily{phv}\selectfont}
\usetikzlibrary{math}
\usetikzlibrary{backgrounds}
\usetikzlibrary{calc,patterns,decorations.pathmorphing,decorations.markings}
\usetikzlibrary{arrows, arrows.meta}
\tikzset{
    partial ellipse/.style args={#1:#2:#3}{
        insert path={+ (#1:#3) arc (#1:#2:#3)}
    }
}

\begin{document}
\fphead
\hspace*{-2.5mm}\begin{tabular}{||p{\textwidth}}
\begin{center}
\vspace{-4mm}
\title{EURODINAME-2026-12929\\
LEARNING-BASED CONTROL OF A SINGLE-DOF AERO SYSTEM} 
\end{center}
\authors{Gabriel da Silva Lima} \\
\authors{Wallace Moreira Bessa} \\
\institution{Smart Systems Lab, Department of Mechanical Engineering, University of Turku, 20520, Finland.} \\
\institution{gdasil@utu.fi, wmobes@utu.fi} \\
\\
\abstract{\textbf{Abstract.} This paper presents a learning-based control framework that integrates feedback linearization with reinforcement learning for the adaptive control of nonlinear mechatronic systems. The control law is derived using Lyapunov stability analysis, ensuring closed-loop stability in the presence of modeling uncertainties and external disturbances. Feedback linearization serves as the main control framework, while a reinforcement learning component estimates and compensates for unmodeled dynamics and disturbances online. The learning module is based on the REINFORCE-with-baseline algorithm, which improves learning efficiency by reducing the variance of policy-gradient estimates and enabling stable policy updates during adaptation. The proposed controller is evaluated on a single-degree-of-freedom rotor-based AERO system. Results from simulations demonstrate accurate trajectory tracking, fast adaptation, and strong robustness against parameter variations and external disturbances. Overall, the proposed approach combines the analytical guarantees of Lyapunov-based control with the adaptability of reinforcement learning, providing an effective solution for controlling nonlinear mechatronic systems.}\\

\\
\\
\end{tabular}

\section{INTRODUCTION}

Nonlinear mechatronic systems are ubiquitous in modern engineering applications, including aerospace platforms, robotic manipulators, and electromechanical devices. The control of such systems is particularly challenging due to strong nonlinearities, parametric uncertainties, unmodeled dynamics, and external disturbances, which often degrade performance and compromise stability when conventional model-based controllers are applied~\citep{da2023accurate, misganaw2025development}. Classical nonlinear control techniques, such as feedback linearization and sliding mode control, offer powerful tools to handle nonlinear dynamics and provide formal stability guarantees; however, their performance critically depends on the availability of accurate system models~\citep{turab2025adaptive, ismael2026robust}. In practice, modeling errors and unknown disturbances are unavoidable, motivating the development of adaptive, intelligent, and learning-based control strategies that can preserve stability while improving robustness and performance in uncertain environments.

While various adaptive and learning-based control strategies have been proposed to address modeling inaccuracies and disturbances, many approaches suffer from limitations related to convergence speed, sample efficiency, or lack of formal stability guarantees~\citep{berkenkamp2017safe, cao2025reinforcement}. Reinforcement learning (RL), in particular, has emerged as a powerful framework for adaptive control due to its capacity to learn optimal policies through interaction with the environment~\citep{sutton1998reinforcement}. However, standard RL algorithms often require extensive exploration and generally do not provide formal guarantees of closed-loop stability, particularly when applied independently to safety-critical or highly nonlinear systems~\citep{chow2018lyapunov, zhang2024model}. These challenges have motivated the development of hybrid control strategies that combine the stability guarantees of model-based control techniques with the adaptability of learning-based components. In such frameworks, the model-based controller provides a stable foundation, while the learning module adapts to unmodeled dynamics and uncertainties. However, most RL-based controllers continue to be developed independently of classical control theory, which can limit their interpretability and complicate deployment in real-world applications. This disconnect underscores the need for integrated control architectures that leverage the complementary strengths of both paradigms, ensuring stability through analytical design while enabling robustness and adaptability through reinforcement learning.

In this work, we develop and evaluate a hybrid control strategy for a single-degree-of-freedom (DoF) model of the Quanser AERO system, a laboratory-scale aerospace platform commonly used in control education and research. The controller combines feedback linearization with a policy-gradient reinforcement learning algorithm to achieve robust trajectory tracking in the presence of model uncertainties and external disturbances. A Lyapunov-based nominal controller ensures closed-loop stability, while an actor–critic reinforcement learning agent provides online compensation for unmodeled dynamics. The reinforcement learning component is implemented using the REINFORCE-with-baseline algorithm, which incorporates advantage estimation to improve learning efficiency and policy stability. Numerical simulations demonstrate the feasibility and effectiveness of the proposed architecture in adapting to dynamic uncertainties while maintaining stable control.

\section{SYSTEM MODEL}

\begin{figure}[ht]
    \centering
    \resizebox{0.8\textwidth}{!}{\input{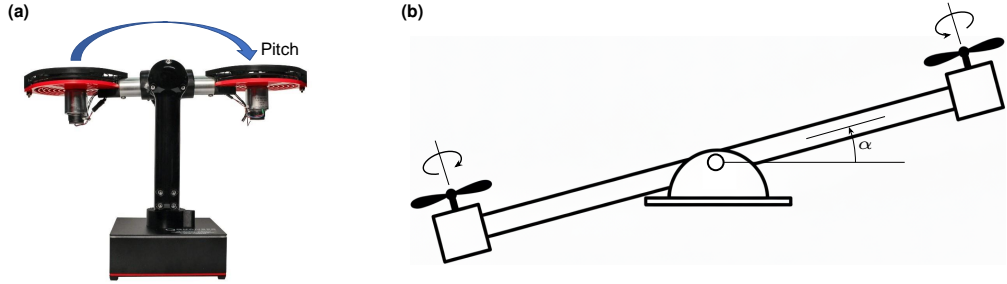}}
    \caption{Single-DOF AERO system: (a) Quanser device; (b) simplified representation.}
    \label{fig:illust}
\end{figure}

The Quanser AERO 2, Fig.~\ref{fig:illust}(a), is a dual-rotor aerospace laboratory experiment used in engineering education and research to teach and test control systems and dynamics concepts.  It allows users to work with a compact physical setup that captures the essential challenges of flight control in a safe, laboratory-scale environment. Although the system has two degrees of freedom (DoF), pitch and yaw, for simplicity purposes, in this work we will work with a reduced model with only one DoF, pitch. A simplified representation of the system is shown in Fig.~\ref{fig:illust}(b). The corresponding dynamic model is given by Eq.~(\ref{eq:model}).
\begin{equation}\label{eq:model}
    I_p\ddot{\alpha} + M_b g D_m \sin\alpha = 2 K_v D_t V_m
\end{equation}
with $\alpha$ being the pitch angle, $I_{p}$ the moment of inertia, $M_b$ the mass of the body, $g = 9.81$~m/s$^2$ the gravitational acceleration, $D_m$ the distance between the pitch axes and the center of mass, $K_v$ the proportionality constant to convert motor voltage to thrust force, $D_t$ the distance between the rotor center and the pitch axes, and $V_m$ the motor voltage. It should be noted that, to avoid zero net momentum, we apply voltages with the same absolute value to both propellers but with opposite signs, which results in thrust forces acting in opposite directions.

\section{LEARNING-BASED CONTROLLER}

In order to derive the control law, the dynamic model will be rewritten as
\begin{equation}\label{eq:model2}
    \ddot{\alpha} = f + b V_m
\end{equation}
where $f$ is a nonlinear function that represents the system dynamics and $b$ is the input gain. Considering that the system is subjected to modeling uncertainties, i.e., $f = \hat{f} + \Delta f$ and $b = \hat{b} + \Delta b$, Eq.~(\ref{eq:model2}) can reformulated as
\begin{equation}\label{eq:model_final}
    \ddot{\alpha} = \hat{f} + \hat{b} V_m + d
\end{equation}
where $d$ comprises the modeling uncertainties $\Delta f$ and $\Delta b$.

A combined error signal inspired by the sliding mode method is defined as $\eta = \dot{\tilde{\alpha}} + \lambda \tilde{\alpha}$~\citep{slotine1991applied}, where $\lambda$ is a strictly positive constant, $\tilde{\alpha} = \alpha - \alpha_d$ is the tracking error, and $\alpha_d$ denotes the desired angle. Following the feedback linearization approach, the control law for the system represented by Eq.~(\ref{eq:model_final}) is given by
\begin{equation}\label{eq:control_law}
    V_m = \hat{b}^{-1} (- \hat{f} - \hat{d} + \ddot{\alpha}_d - \lambda \dot{\tilde{\alpha}} - \lambda \eta)
\end{equation}

Applying the control law~(\ref{eq:control_law}) into Eq.~(\ref{eq:model_final}) yields
\begin{equation}\label{eq:closed_loop}
    \dot{\eta} + \lambda \eta = \tilde{d}
\end{equation}
where $\tilde{d} = d - \hat{d}$ is the approximation error. Assuming that we have a perfect estimation of the system dynamics, i.e. $\hat{d} = d$, then the combined error $\eta$, and consequently the tracking error $\tilde{\alpha}$, will converge to zero. Otherwise, we can observe that the closed-loop dynamics are governed by the approximation error $\tilde{d}$. 

\subsection{Reinforcement Learning Compensation}

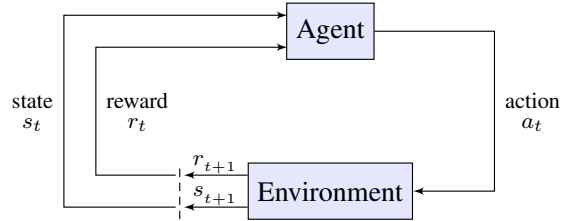
\begin{figure}[ht]
    \centering
    \resizebox{0.5\textwidth}{!}{\tikzstyle{box} = [draw, fill=blue!10, rectangle, minimum height=2em, minimum width=2em]
\tikzstyle{block} = [draw, fill=blue!10, rectangle, minimum height=2em, minimum width=7em]
\tikzstyle{sum} = [draw, fill=blue!10, circle, node distance=1cm]
\tikzstyle{place} = [coordinate]
\tikzstyle{pinstyle} = [pin edge={to-,thin,black}]
\begin{tikzpicture}[auto, node distance=2cm,>=latex']

    \node at (0,-1) [box, align=center] (E) {Environment};

    \node at (0,1) [box, align=center] (A) {Agent};

    \draw [->] (A.east) -- ++(1.5,0) -- node[right=-2mm] {\begin{tabular}{c}
                                                                                \footnotesize action\\[-5pt]
                                                                                \footnotesize $a_t$
                                                                            \end{tabular}} ++(0,-2) -- (E.east);

    \draw[->] (E.west) ++(0,0.2) -- node[midway, above=-1mm] {{\footnotesize$r_{{}_{t+1}}$}} ++(-0.8,0);
    \draw[->] (E.west) ++(0,-0.2) -- node[midway, above=-1mm] {{\footnotesize$s_{{}_{t+1}}$}} ++(-0.8,0);

    \draw[densely dashed] ($(E.west)+(-0.85,-0.35)$) -- ++(0,0.7);

    \draw[->] ($(E.west)+(-0.9,0.2)$) -- ++(-1.0,0) -- node[right=-2mm] {\begin{tabular}{c}
                                                                                \footnotesize reward\\[-5pt]
                                                                                \footnotesize $r_t$
                                                                            \end{tabular}} ++(0.0,1.6) -- ++(2.4,0.0);
    \draw[->] ($(E.west)+(-0.9,-0.2)$) -- ++(-1.4,0) -- node[left=-2mm] {\begin{tabular}{c}
                                                                                \footnotesize state\\[-5pt]
                                                                                \footnotesize $s_t$
                                                                            \end{tabular}} ++(0.0,2.4) -- ++(2.8,0.0);




	
\end{tikzpicture}}
    \caption{Reinforcement learning framework.}
    \label{fig:RL}
\end{figure}

The reinforcement learning process involves an agent that continuously interacts with the environment by selecting actions based on its current state and receiving a feedback signal known as a reward that evaluates the chosen action~\citep{sutton1998reinforcement}, as illustrated in Fig.~\ref{fig:RL}.

At each discrete time $t$, the agent observes the system state $s_t \in \mathcal{S}$, selects an action $a_t \in \mathcal{A}$ according to a decision rule or policy $\pi_{\bm{\theta}}$, where $\bm{\theta} \in \mathbb{R}^q$ is the policy's parameter vector, and the system evolves to a new state $s_{t+1}$ while producing a scalar reward $r_{t+1} \in \mathbb{R}$. This interaction is commonly formalized as a Markov Decision Process (MDP), characterized by the tuple $(\mathcal{S}, \mathcal{A}, \mathcal{P}, r, \gamma)$, where $\mathcal{P}(s_{t+1} \vert s_t, a_t)$ denotes the state-transition probability, $r(s_t, a_t)$ is the reward function, and $\gamma \in (0,1]$ is the discount factor that determines the relative importance of future rewards. The objective of the agent is to learn a policy $\pi_{\bm{\theta}}(a_t \vert s_t)$ that maximizes the expected cumulative discounted return
\begin{equation}\label{eq:return}
    J(\bm{\theta}) = \mathbb{E}_{s_0 \thicksim \rho_0, \pi_{\bm{\theta}}} \left[ \sum_{t=1}^T \gamma^{t-1} r_t \right]
\end{equation}
where $\rho_0$ denotes the distribution of initial states $s_0$.

RL algorithms are generally categorized into value-based and policy-based methods. While value-based methods attempt to estimate the utility of being in a state, policy-based methods directly parameterize the policy $\pi_{\bm{\theta}}$ and optimize the parameters $\bm{\theta}$ using gradient ascent. This approach is particularly advantageous in continuous action spaces, such as those found in control tasks, where searching for a greedy action over a value function becomes computationally expensive. Furthermore, policy gradients can naturally learn stochastic policies, which provide a built-in mechanism for exploration and are robust to the perceptual aliasing often encountered in real-world environments.

To optimize the objective function defined in Eq.~(\ref{eq:return}), the REINFORCE algorithm employs the Policy Gradient Theorem~\citep{sutton1999policy} to estimate the gradient of the expected return. In this work, the policy (actor) is modeled as a Gaussian distribution $\mathcal{N}(\mu_{\bm{\theta}}, \sigma^2)$, where the mean $\mu_{\bm{\theta}}$ is the output of a neural network and the standard deviation $\sigma$ is treated as a learnable parameter and included in the parameter vector $\bm{\theta}$. By sampling an action $a_t$ from this distribution, the agent explores the environment and collects a trajectory of rewards. The parameters $\bm{\theta}$ are then updated in the direction of the gradient:
\begin{equation}\label{eq:gradient}
    \nabla_{\bm{\theta}} J(\bm{\theta}) = - \mathbb{E}_{\pi_{\bm{\theta}}} \left[ G_t \nabla_{\bm{\theta}} \ln \pi_{\bm{\theta}}(a_t \vert s_t) \right]
\end{equation}
where $G_t = \sum_{k=t}^T \gamma^{k-t} r_k$ represents the discounted return from time $t$.

To reduce the high variance associated with the vanilla REINFORCE~\citep{williams1992simple}, a state-dependent baseline is introduced~\citep{sutton1999policy}. In this architecture, the baseline is represented by a separate neural network, known as the critic, which provides a functional approximation $V^{\pi_{\bm{\theta}}} (s_t)$ of the state-value function:
\begin{equation}\label{eq:state_value}
    V^{\pi_{\bm{\theta}}} (s_t) = \mathbb{E}_{\pi_{\bm{\theta}}} \left[ \sum_{k=t}^T \gamma^{k-t} r_k \mid s_t \right]
\end{equation}

To further stabilize training, \cite{schulman2015high}~proposed an advantage-based update by employing the Generalized Advantage Estimation (GAE). The temporal difference $\delta_t$ and the resulting advantage $A_t$ are calculated as:
\begin{equation}\label{eq:TD}
    \delta_t = r_t + \gamma V(s_{t+1}) - V(s_t)
\end{equation}
\vspace{-12pt}
\begin{equation}\label{eq:advantage}
    A_t = \delta_t + \gamma \beta A_{t+1}
\end{equation}
where $\beta \in [0,1]$ is a smoothing parameter. The advantage is normalized across the batch to reduce gradient variance and ensure consistent step sizes during optimization.

In this dual-network framework, the actor and critic are trained by minimizing the actor and critic losses. In this case, the gradient function~(\ref{eq:gradient}) is modified to take into account the advantage instead of the discounted return:
\begin{equation}\label{eq:actor}
    \mathcal{L}_{actor} = - \sum_{t=1}^T  A_t \ln \pi_{\bm{\theta}}(a_t \vert s_t)
\end{equation}
while the critic loss is the Mean Squared Error (MSE) of the value predictions:
\begin{equation}\label{eq:critic}
    \mathcal{L}_{critic} = \frac{1}{T} \sum_{t=1}^{T} (V^{\pi_{\bm{\theta}}}(s_t) - G_t)^2
\end{equation}
By employing the normalized advantage to drive policy updates, the agent ensures that parameters are adjusted based on the relative benefit of an action compared to the state-value baseline, promoting more stable convergence in continuous control tasks.

In our control application, the action $a_t$ taken by the agent corresponds to the disturbance compensator $\hat{d}$ at time step $t$. During the training step, in order to encourage exploration, $\hat{d}_t$ is obtained by sampling from the stochastic policy parameterized by the actor neural network, given by
\begin{equation}\label{eq:d_hat_train}
    \hat{d}_t \thicksim \mathcal{N}(\mu_{\bm{\theta}}(s_t), \sigma^2)
\end{equation}

During deployment, the disturbance compensator is chosen deterministically as
\begin{equation}\label{eq:d_hat_deploy}
    \hat{d}_t = \mu_{\bm{\theta}}(s_t)
\end{equation}
%

%
%
%
%

To finalize the reinforcement learning algorithm, the reward function is defined based on the combined error $\eta$. Recalling the closed-loop dynamics given in~(\ref{eq:closed_loop}), the disturbance approximation error directly affects the evolution of $\eta$. Consequently, reducing the magnitude of $\eta$ leads to an improved approximation of the unknown disturbance.

The reward is therefore defined as
\begin{equation}\label{eq:reward}
    r = -|\eta| \ln(1 + t)
\end{equation}
where the logarithmic time-dependent term attenuates the influence of large transient errors at the beginning of each episode, allowing the learning process to focus progressively on steady-state performance.

\subsection{Stability Analysis}

Before analyzing the stability properties of the proposed control law, we first state the assumptions required for the stability proof:

\noindent
\textit{Assumption 1.} The pitch angle $\alpha$ and its derivative are measurable or can be estimated.

\noindent
\textit{Assumption 2.} The desired trajectory $\alpha_d$ is available and is a continuously differentiable function.

\noindent
\textit{Assumption 3.} The disturbance term is unknown but bounded, i.e. $|d| \leq \xi$.

Following a Lyapunov-type stability analysis, let a positive-definite function $V(t)$ be defined as:
\begin{equation}
    V(t) = \frac{1}{2} \eta^2
\end{equation}

Taking the first derivative of $V(t)$, we have:
\begin{align}
    \dot{V}(t) & = \eta \dot{\eta} \nonumber \\
               & = \eta (\hat{f} + \hat{b} V_m + d - \ddot{\alpha}_d + \lambda\dot{\tilde{\alpha}}) \nonumber \\
               & = \eta (\tilde{d} - \lambda\eta) \label{eq:Vdot}
\end{align}

From the closed-loop dynamics in~(\ref{eq:closed_loop}), it can be observed that the convergence of the combined error variable $\eta$ is directly affected by the disturbance approximation error $\tilde{d}$. By defining a reward function that encourages the reinforcement learning algorithm to reduce the magnitude of $|\eta|$, and assuming that the output of the RL agent is bounded, it follows that the disturbance approximation error remains bounded, i.e., $|\tilde{d}| < \varepsilon$. Consequently, as the REINFORCE-based policy converges to a locally optimal solution of the disturbance approximation problem, the closed-loop system exhibits bounded tracking errors.

Hence, $\dot{V}(t)$ becomes
\begin{equation}
    \dot{V}(t) \leq - [\lambda|\eta| - \varepsilon] |\eta|
\end{equation}
Therefore, $\dot{V}(t)$ is negative-definite for $|\eta| \geq \varepsilon/\lambda$. For $|\eta| < \varepsilon/\lambda$, recalling Assumption~3, and the boundedness of the disturbance estimate $\hat{d}$, it follows that the combined error $|\eta|$ is uniformly ultimately bounded within a compact set of radius $\varepsilon/\lambda$. Consequently, the closed-loop signals $|\tilde{\alpha}|$ and $|\dot{\tilde{\alpha}}|$ are also uniformly ultimately bounded within the same region~\citep{bessa2009some}. It is important to note that closed-loop stability is ensured by the Lyapunov-based feedback linearization controller, while the learning agent contributes with bounded adaptive compensation without violating the system’s stability conditions.

\section{NUMERICAL RESULTS}

Numerical simulations were conducted in Python by applying the proposed RL and control schemes to the AERO system described in Eq.~(\ref{eq:model}). The fourth-order Runge-Kutta method was used for numerical integration, with sampling rates of 10 Hz and 100 Hz for the controller and system dynamics, respectivelly. Table~\ref{tab:parameters} presents the selected values for the dynamic model and controller paramters. The control gain was set to $\lambda = 3$. 

For the RL algorithm, the discount factor was set to $\gamma = 0.999$ and the GAE smoothing parameter to $\beta = 0.98$. Both the actor and critic were modeled as fully connected neural networks with architecture $64 \times 64 \times 64$. To ensure bounded action outputs during training and deployment, the actor’s mean output was constrained within the range [-10, 10], ensuring compatibility with the assumptions in the Lyapunov-based stability analysis, and the log standard deviation was parameterized as a learnable variable initialized to zero. The system state (input) was defined as the tuple $s = (\alpha, \dot{\alpha}, \alpha_d, \dot{\alpha}_d)$. 


Training was performed over a maximum of 20,000 episodes, with each episode comprising up to 60\,s of simulation. The policy was updated using the Adam optimizer with different initial learning rates ($\kappa$), which decayed linearly to zero over the training horizon. Gradient clipping with a maximum norm of 0.5 was applied to ensure numerical stability. During training, the initial system conditions were sampled from a uniform distribution: $\rho_0 = \{ \alpha(0) \sim \mathcal{U}(\pi/6, \pi/3), \dot{\alpha}(0) \sim \mathcal{U}(-0.1, 0.1) \}$. The desired trajectory was defined as $\alpha_d(t) = \pi/4 \sin\left(\pi t/5\right)$, where $\pi$ denotes the mathematical constant $\pi \approx 3.14159$.

Early stopping was implemented based on the Wasserstein distance between return distributions over sliding windows of 400 episodes, with a patience of 400 episodes and a minimum improvement threshold $\delta_\text{mean} = 500$. During training, the desired trajectory's amplitude and frequency were randomized every 400 episodes around their nominal values to introduce nonstationarity. Evaluation was performed in simulation using a fixed seed and without stochastic exploration, applying the mean of the learned policy as the deterministic control action.
\begin{table}[ht]
    \centering
    \begin{tabular}{l|cc}
       Parameter & Dynamic model  & Controller \\ \hline
       $I_p$ (kg.m$^2$) & $2.15 \times 10^{-2}$ & $10^{-2}$ \\
       $M_b$ (kg) & 1.075 & 1\\
       $D_m$ (mm) & 7.95 & 5\\
       $K_v$ (mN/V) & 5 & 3\\
       $D_t$ (mm) & 158 & 200
    \end{tabular}
    \caption{Model and controller parameters.}
    \label{tab:parameters}
\end{table}
%

\begin{figure}[ht]
    \centering
    \resizebox{0.9\textwidth}{!}{\input{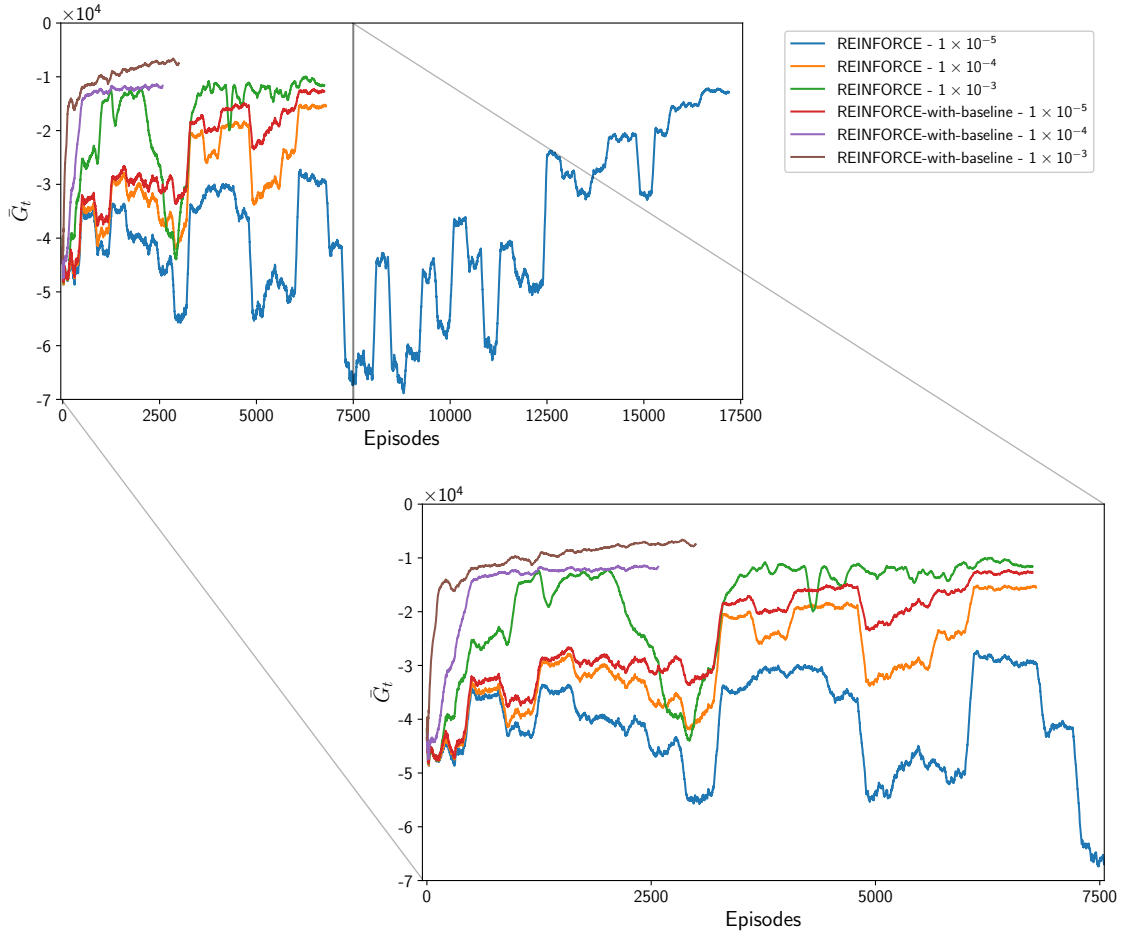}}
    \caption{Average return $\bar{G}_t$ over episodes for REINFORCE and REINFORCE-with-baseline\\ using three different learning rates.}
    \label{fig:return}
\end{figure}

Figure~\ref{fig:return} shows the evolution of the average return $\bar{G}_t$ over training episodes for REINFORCE and REINFORCE-with-baseline using three different learning rates. Each curve represents the smoothed return from a single training run, with early stopping triggered when return improvement plateaued over a 400-episode window. Across all learning rates, the REINFORCE-with-baseline algorithm consistently achieved higher final returns and faster convergence compared to its vanilla counterpart. Notably, with a learning rate of $1 \times 10^{-4}$, the baseline-enhanced variant reached stable performance in approximately 2,500 episodes and maintained the second-highest return throughout training, suggesting a favorable trade-off between speed and stability.

In contrast, vanilla REINFORCE exhibited significantly more unstable learning behavior, with wider fluctuations and slower progression, particularly at lower learning rates. For example, with $\kappa = 1 \times 10^{-5}$, REINFORCE failed to converge even after 15,000 episodes. The learning rate had a noticeable impact on convergence: higher values generally led to earlier stopping (due to faster learning), but at the risk of instability in the vanilla case. The inset plot magnifies the first 7,500 episodes, highlighting early performance gaps, especially the more monotonic ascent achieved by REINFORCE-with-baseline.

These results underscore the stabilizing effect of baseline estimation in policy-gradient methods, especially in systems with nonstationary dynamics. For the control of nonlinear mechatronic systems, incorporating a baseline not only improves sample efficiency but also leads to smoother and more reliable policy learning under varying conditions. 

\begin{figure}[ht]
    \centering
    \includegraphics[width=\textwidth]{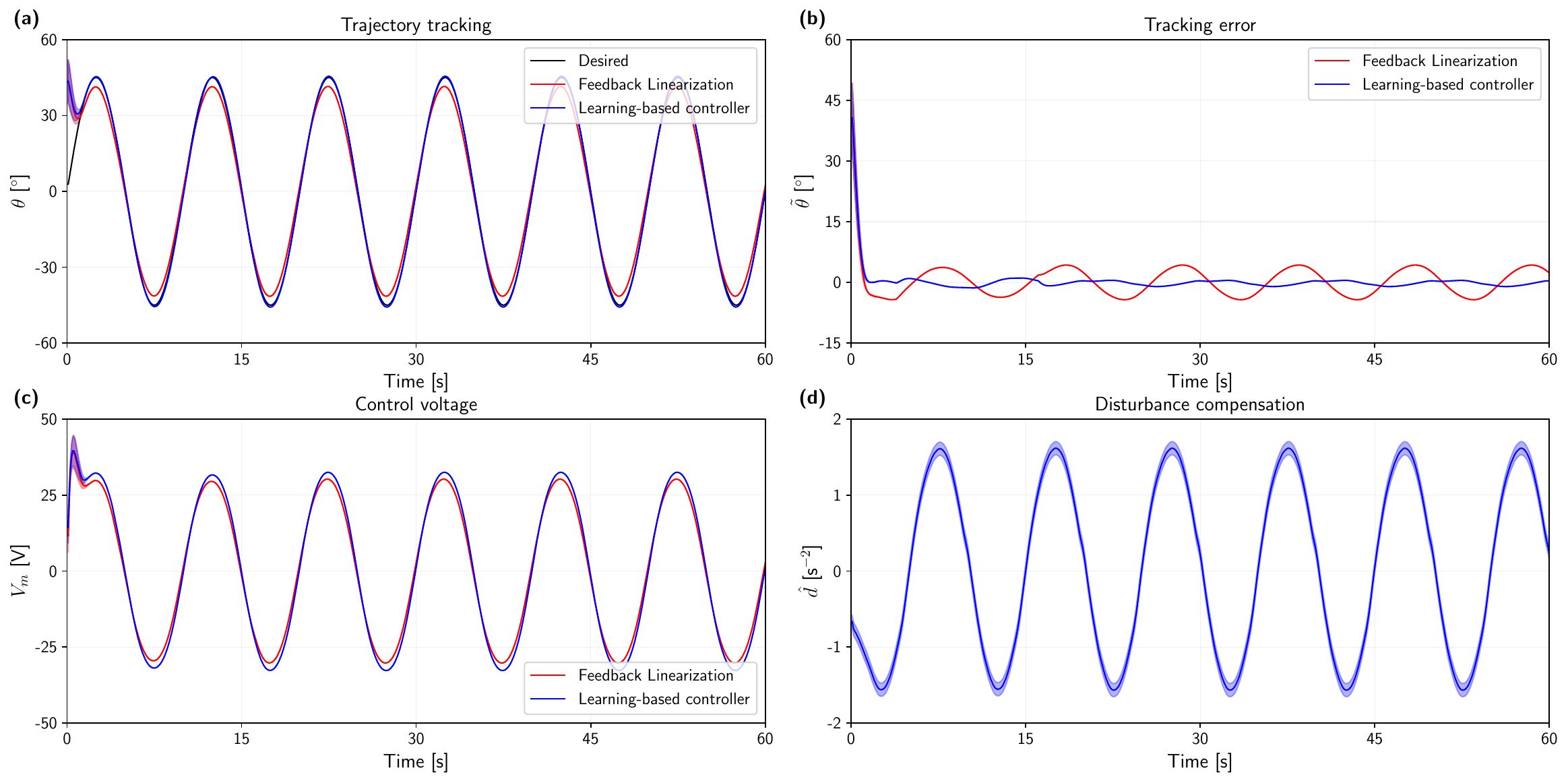}
    \caption{Averaged results over 100 simulations with randomized initial conditions. Comparison between the conventional feedback linearization controller and the proposed learning-enhanced controller, which augments feedback linearization with a disturbance compensator trained via REINFORCE-with-baseline ($\kappa = 1 \times 10^{-4}$).}
    \label{fig:graph1}
\end{figure}

To assess the impact of the proposed controller, we conducted 100 independent simulations with randomized initial conditions. The proposed approach integrates a disturbance compensator, learned via REINFORCE-with-baseline using a learning rate of $1 \times 10^{-4}$, into a feedback linearization framework. We compared its performance against the conventional feedback linearization controller alone, evaluated under identical test conditions. Figure~\ref{fig:graph1} presents the average system behavior across all trials.

The learning-based controller significantly improved trajectory tracking accuracy. As shown in Fig.~\ref{fig:graph1}(a), the proposed control strategy closely follows the desired sinusoidal reference, with visibly reduced steady-state and transient deviations compared to the feedback linearization approach. This improvement is further highlighted in subplot (b), where the tracking error $\tilde{\theta}$ is consistently smaller for the learning-based controller, indicating enhanced robustness to initialization variability. Importantly, subplot (c) shows that these gains were achieved without introducing excessive control effort, since we can observe the control voltage profiles remain similar across both methods. Finally, subplot (d) illustrates the learned disturbance compensation $\hat{d}$, which oscillates in synchrony with the system dynamics and contributes to the improved performance.

To evaluate the robustness of the proposed controller under unmodeled time-varying disturbances, we conducted other new 100 simulations where an additional disturbance signal was injected into the system dynamics. This disturbance followed the form $d(t) = A_d \left[\sin(\omega_d t) + \cos(\omega_d t) + 0.2 \sin(\omega_d t)\cos(\omega_d t)\right]$, with amplitude $A_d$ and frequency $\omega_d$ randomly sampled for each run. The same trained REINFORCE-with-baseline policy ($\kappa = 1 \times 10^{-4}$) was reused without retraining or adaptation to the new disturbance profile.

\begin{figure}[ht]
    \centering
    \includegraphics[width=\textwidth]{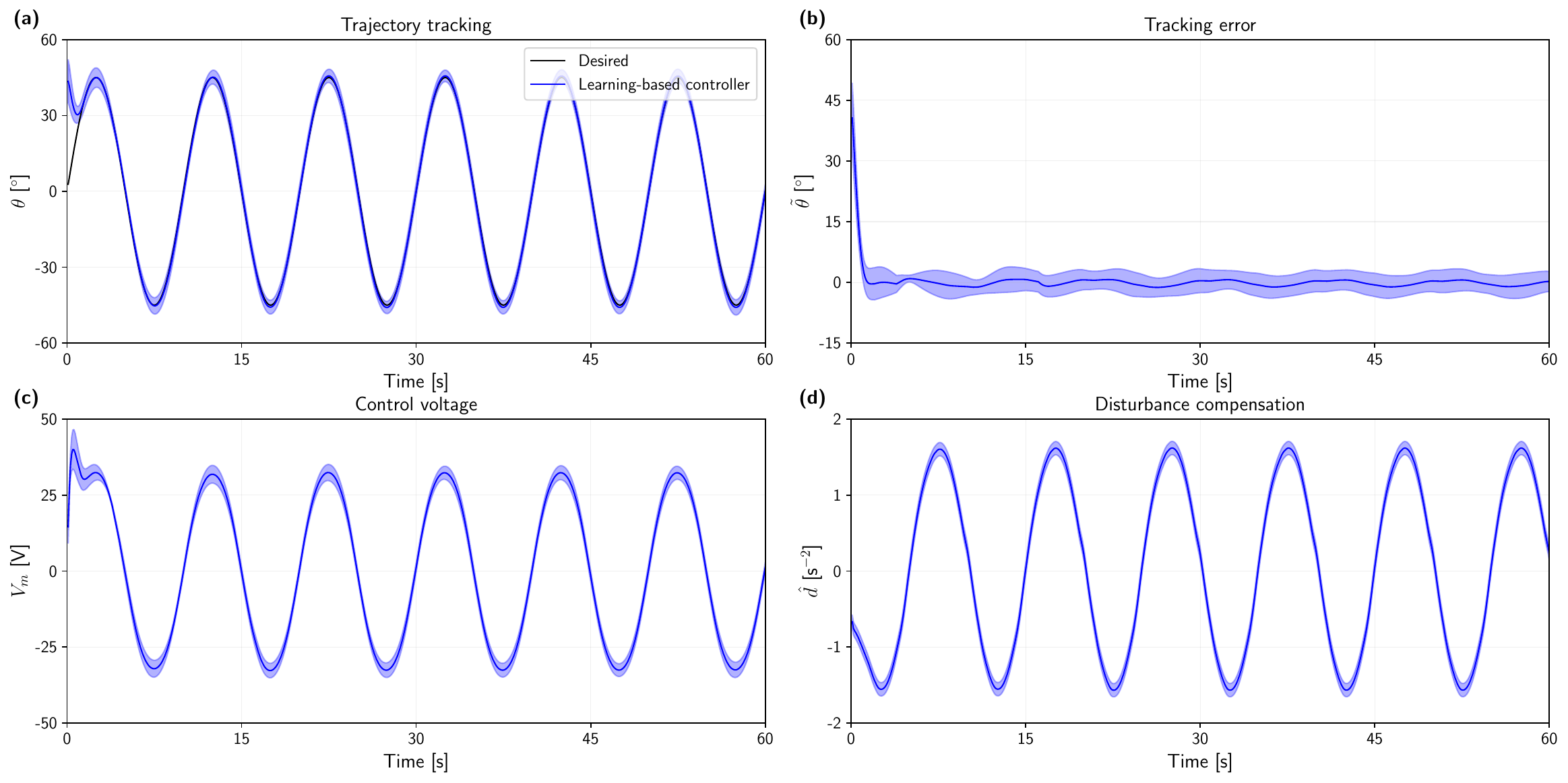}
    \caption{Robustness evaluation under randomly generated external disturbances. Results averaged over 100 simulation runs using the trained REINFORCE-with-baseline controller ($\kappa = 1 \times 10^{-4}$). An additional disturbance signal with randomly sampled amplitude and frequency was applied in each run.}
    \label{fig:graph2}
\end{figure}

Figure~\ref{fig:graph2} presents the averaged system response. Despite the presence of unknown and varying disturbances, the learning-based controller maintained high tracking performance. As shown in Fig.~\ref{fig:graph2}(a), the output trajectory remains tightly aligned with the desired reference, and subplot (b) confirms that the tracking error stays low and bounded across time. Figure~\ref{fig:graph2}(c) shows that the control voltage remains well-regulated, while subplot (d) illustrates the adaptive disturbance compensation learned by the policy, which captures the oscillatory structure of the injected signal. These results indicate that the proposed hybrid controller generalizes well and retains its disturbance rejection capability even in dynamically perturbed environments.

\section{CONCLUDING REMARKS}

This work presented a hybrid learning-based control framework for nonlinear mechatronic systems, combining feedback linearization with reinforcement learning to address unmodeled dynamics and external disturbances. The control structure was derived using a Lyapunov-based design to ensure closed-loop stability, while the reinforcement learning component, in which was implemented via the REINFORCE-with-baseline algorithm, served as an adaptive disturbance compensator. Numerical simulations on a single-degree-of-freedom model of the Quanser AERO system demonstrated the effectiveness of the proposed approach in achieving accurate trajectory tracking and strong robustness to both parametric variations and time-varying disturbances.

Across all tested scenarios, the learning-enhanced controller consistently outperformed the conventional feedback linearization strategy, yielding lower tracking errors without increasing control effort. These results highlight the value of combining model-based control stability guarantees with data-driven adaptability to improve performance in uncertain control environments. Although the reinforcement learning agent does not independently ensure stability, it operates within the bounds of the conventional controller and enhances system performance by adaptively estimating unmodeled effects. This clear separation of responsibilities enables safe and stable learning in the presence of uncertainty.


\section{REFERENCES} 

\bibliographystyle{eurodiname2026}
\renewcommand{\refname}{}
\bibliography{bibfile}

\section{RESPONSIBILITY NOTICE}

The authors are solely responsible for the printed material included in this paper.

\end{document}